\documentclass[]{spie}  
\usepackage[]{amsmath}
\usepackage[]{graphicx}

 \newcommand{\V}[1]{\boldsymbol{#1}}      
 \newcommand{\M}[1]{\mathbf{#1}}          
 \newcommand{\T}{^\mathrm{T}}             

\newcommand{\Norm}[1]{\left\Vert #1\right\Vert}

\newcommand{\Avg}[1]{\left\langle #1\right\rangle}

\usepackage{xspace}
\newcommand{\ie}{\emph{i.e.}\xspace}     

\title{Optimization of adaptive optics correction \\ during observations:\\
  Algorithms and system parameters identification in closed-loop}

\author{Cl\'ementine B\'echet, Michel Tallon and \'Eric Thi\'ebaut
  \skiplinehalf
  Universit\'e Lyon 1, Villeurbanne, F-69622, France\\
  Centre de Recherche Astrophysique de Lyon, Observatoire de Lyon, 9 avenue Charles Andr\'e, Saint-Genis Laval cedex, F-69561, France\\
  CNRS, UMR 5574; Ecole Normale Sup\'erieure de Lyon, Lyon, France}

\authorinfo{Further author information: (Send correspondence to
  Cl\'ementine B\'echet)\\E-mail: Clementine.Bechet@univ-lyon1.fr,
  Telephone: +33 (0) 4 78 86 85 37}

\begin{document} 
  \maketitle 

\begin{abstract}
  The adaptive optics (AO) on the European Extremely Large Telescope,
  as well as earlier pathfinders like the Adaptive Optics Facility, at
  the Very Large Telescope in 2014, will no longer be stationary
  systems. AO is no longer isolated on a bench; some elements are
  directly in the optical train of the telescope, suffering
  environment and constrains changes during the observations. To
  guarantee good performance at any observing time, we investigate a
  self-calibration strategy. We focus here on one of the most
  challenging aspects: the identification of system parameters during
  closed-loop observations without introducing any additional
  disturbance. Such problem is known in the identification theory to
  be difficult to solve. We have recently presented (B\'echet
  \textit{et al.}, AO4ELT2 Conference, 2011) an identification method
  for this, with promising results obtained in simulations. To
  consolidate these advances, we come back in the present paper to the
  equations and provide a theoretical analysis to justify the choice
  of the algorithm. We highlight the benefit of using incremental data
  and commands to decorrelate the disturbance. We also present 2
  implementations of the method, currently studied at the European
  Southern Observatory.
\end{abstract}


\keywords{identification, optimization, adaptive optics, parameters
  estimation, closed-loop, telemetry, maximum likelihood}

\section{INTRODUCTION}
\label{sec:intro}  

Current adaptive optics (AO) usually work fine after a calibration of
the system which remains valid during a long-exposure or even several
nights of observations. This is the case when the AO components, like
the deformable mirrors and the wavefront sensors, are gathered on an
isolated AO bench. Nevertheless, a new generation of AO system appears
which require much more frequent calibrations: the AO systems of the
\textit{adaptive telescopes}. By adaptive telescope, one means a
telescope in which a deformable mirror is directly settled in the
optical train of the telescope, as for instance when a deformable
secondary mirror (DSM) is used.

The Multi Mirror Telescope (MMT) and the Large Binocular Telescope
(LBT) are two telescopes already observing with such a deformable
secondary mirror \cite{EspositoRiccardi2010a}. The European Extremely
Large Telescope (E-ELT), designed by the European Southern Observatory
(ESO), is also expected to be an adaptive telescope, with its fourth
mirror being a deformable mirror for the AO. And before that, ESO will
install the Adaptive Optics Facility (AOF), which includes a DSM, in
2014 at the Very Large Telescope (VLT) in Chile
\cite{ArsenaultMadec2010a}. The VLT-AOF design, without intermediate
focus, will prevent calibrating the interaction matrix (IM) of the AO
systems using an internal reference source. This leads ESO to develop
new calibration strategies, such as the use of a pseudo-synthetic IM
and on-sky calibration of its long-term fixed parameters. A report on
these preparatory activities is presented in Kolb \textit{et al.}
\cite{KolbEtAl2012}.

The VLT-AOF design is a key feature of the GALACSI AO system feeding
the multi Unit Spectroscopic Explorer (MUSE), a panoramic integral
field spectrograph for the detection of young galaxies in the
visible. To reach such a goal, MUSE \cite{LoupiasBacon2010a} and
GALACSI \cite{ArsenaultMadec2010a} will accumulate exposures up to
hundred of hours, these exposures being cut into one-hour
shots. Reaching 5 to 10\% of Strehl in the visible with GALACSI for
MUSE during one-hour will require a very stable performance of the AO
correction. Therefore, in addition to efficient on-sky calibration
strategies, we need a constant optimization of the AO interaction
matrix in order to track changes in the system parameters. For
instance, misregistrations between the deformable mirror and the
wavefront sensors of the AO may progressively appear during exposures,
induced by gravity or temperature variations. The present paper
studies an on-sky identification method of AO interaction matrix
parameters during closed-loop correction. To track the changes of the
systems, such estimation of parameters needs to be done with
approximately a minute-scale period.

The need for identification of interaction matrix parameters, like
shifts or rotation misregistrations, has been already enhanced thanks
to simulations of AOF-like systems \cite{BechetTallon2012}. This study
highlighted loss of performance or even instabilities when shifts
misregistrations were beyond $\sim$ 15-20\% of a subaperture size
\cite{BechetTallon2012}. A method for identification of the
misregistrations parameters such as differential shifts and rotation
between the DSM and the wavefront sensor has been presented and
successfully used in simulations of a simplified AO system
\cite{BechetTallon2012}. However, the accuracy of this estimation method
has not been clearly analyzed so far. Its accuracy, its convergence
speed, its robustness and its computational complexity need to be
addressed. These characteristics may depend on the AO signal-to-noise
ratio, on the turbulence conditions (\textit{e.g.} Fried parameter $r_0$,
coherence time $\tau_0$, wind specific directions), on the length of the
telemetry data used for estimation, on the AO configuration
(\textit{e.g.} Ground Layer correction \textit{GLAO}, Laser Tomography
AO \textit{LTAO}) and even on the reconstruction and control methods. 

The aim of the current paper is to address these dependencies and
their impact from a theoretical analysis of the estimation
method. This is important because such identification method will be
efficiently used on the real AO systems if and only if we understand
how it works, when it does not work, and how well it can improve the
AO performance. In that purpose, the Section~\ref{sec:EqAndNotations}
reminds the equations and the notations used to represent the
closed-loop system of the AOF. In Section~\ref{sec:MeasEq}, we
introduce a first approach for parameters identification based on the
measurement equation of the closed-loop system. A theoretical and
numerical analysis enhances the strong correlation of the residuals in
this approach, and so its difficult use in practice. Therefore, in
Sect.~\ref{sec:DiffMeasEq}, we introduce the criterion already used
for identification in B\'echet \textit{et al.}
\cite{BechetTallon2012}, based on increments of the measurement
equation. Again, a theoretical and numerical analysis shows the
benefit of this approach to obtain better decorrelation of the
residuals. Finally, in Sect.~\ref{sec:Implem}, two different
implementations of this method based on increment of measurements are
discussed.

\section{AO equations}
\label{sec:EqAndNotations}

The closed-loop AO system is modeled by the set of two equations
\begin{eqnarray}
  \label{eq:MeasEq}
\V{d}_k & = & \M{S} (w_k) - \M{G} \cdot \V{a}_k + \V{e}_k \\
\label{eq:ControlEq}
  \V{a}_{k+\tau} & = & \alpha \, \V{a}_{k+\tau -1} + \beta \, \V{a}_{k+\tau -2} + \gamma \, \M{C}\cdot \V{d}_k 
\end{eqnarray}
where 
\begin{itemize}
\item $\V{d}_k$ is the measurement vector (Shack-Hartmann slopes coming
  from $n_s$ guide stars) at $k-$th loop
\item $w_k$ is a continuous representation of the turbulence in the
  atmosphere volume (tomography is possible) during measurement at $k-$th
  loop
\item $\V{a}_k$ is the DM command vector applied during measurement at
  $k-$th loop
\item $\V{e}_k$ is the noise vector on measurement at $k-$th loop
\item $\M{S}$ is a linear model of the multi guide stars
  Shack-Hartmann sensor, from a continuous representation of the
  turbulence to a discrete set of data
\item $\M{G}$ is the command-to-measurement linear model, \ie interaction matrix
\item $\M{C}$ is the command matrix 
\item $\alpha$, $\beta$ and $\gamma$
  are the scalar parameters of the control law, \ie in case of a pure
  integrator law : $\alpha=1$, $\beta=0$, and $\gamma$ is the gain
\item $\tau$ is the AO delay (number of frames)
\end{itemize}
Note that index $k$ for discrete time vectors is not chosen as
commonly used in AO, but this aims at simplifying equations notations
below. Hence, a measurement vector $\V{d}_k$ has been influenced by
the simultaneous application of the command vector $\V{a}_k$ to the
DM.

The aim of the identification method investigated here is to estimate
parameters on which $\M{G}$, the interaction matrix, depends, using a
sequence of data $\V{d}_k$ and commands $\V{a}_k$ recorded during
closed-loop AO, \ie from the AO telemetry. If $\M{G}$ can be
determined using a model, that is to say a synthetic matrix, only a
few parameters have to be identified, to fully determine the matrix
$\M{G}$. We could also consider that all the matrix coefficients of
$\M{G}$ are parameters to be identified. To keep general notations,
and allow both considerations, we denote the interaction matrix as
$\M{G}(\V{p})$, that is to say as a function of a vector of parameters
$\V{p}$. These are the parameters we want to estimate hereafter. Note
that there is no linearity assumption between the parameters and the
interaction matrix coefficients in our study.

\section{Criterion from the measurement equation}
\label{sec:MeasEq}

The measurement equation~(\ref{eq:MeasEq}) can also be written
\begin{equation}
  \label{eq:MeasEqZ}
  \V{d}_k = -\M{G}(\V{p})\cdot \V{a}_k + \V{z}_k
\end{equation}
where $\V{z}_k=\M{S}(w_k) + \V{e}_k$ is now considered as a disturbance vector,
including both turbulence contribution to the measurements and measurement
noise. From this new point of view, in Eq.~(\ref{eq:MeasEqZ}), the data
$\V{d}_k$ are considered to be noisy measurements of the slopes induced by
the command vector $\V{a}_k$ applied to the DM.

The disturbance, $\V{z}_k$, follows zero-mean Gaussian statistics and its
covariance matrix can be written
\begin{equation}
  \label{eq:CovZkDev}
  \M{C_{z_k}}  = \Avg{\V{z}_k \cdot {\V{z}_k}\T } = \Avg{ \V{s}_k \cdot \V{s}_k\T} + \Avg{\V{e}_k \cdot \V{e}_k \T} = \M{C_{s_k}} + \M{C_{e_k}} \\
\end{equation}
where $\V{s}_k= \M{S}(w_k)$ is the noiseless Shack-Hartmann
measurement of the turbulence $w_k$. Turbulence and noise are
considered to be stationary process at the scale of our estimation
study, so that the covariance matrices of $\V{s}_k$, $\V{e}_k$ and
$\V{z}_k$ are independent of $k$, and eventually written $\M{C_{s}}$,
$\M{C_{e}}$ and $\M{C_{z}}$ respectively, so that
Eq.~(\ref{eq:CovZkDev}) becomes
\begin{equation}
  \label{eq:CovZ}
  \M{C_{z}}  =  \M{C_{s}} + \M{C_{e}}\,.
\end{equation}

\subsection{Criterion}
\label{sec:CritMeasEq}

From Eq.~(\ref{eq:MeasEqZ}), a maximum likelihood approach to estimate
the parameters $\V{p}$ of the interaction matrix $\M{G}$ leads to the
minimization of the following criterion
\begin{equation}
  \label{eq:Chi2NoINCR}
  \chi_{1}^2 (\V{p}) = (\V{d}_k + \M{G}(\V{p})\cdot \V{a}_k) \T \cdot \M{C}_{z}^{-1} \cdot (\V{d}_k + \M{G}(\V{p})\cdot \V{a}_k) \,,
\end{equation}
in case of using only one measurement vector $\V{d}_k$. The lower
indice $1$ of the $\chi_{1}^2$ stands for using only one vector of
measurements and commands.

Selecting several sets of measurements separated by a certain number
of frames, $\Delta T$, large enough to avoid correlations between the
associated $\V{z}_k$ and $\V{z}_{k+\Delta T}$, more residuals can be
added in the $\chi^2$ computation, and the criterion can be changed into
\begin{equation}
  \label{eq:Chi2NoINCR-N}
  \chi_{N,\Delta T}^2 (\V{p}) = \sum_{k=1}^{k=N}(\V{d}_{k\, \Delta T} + \M{G}(\V{p})\cdot \V{a}_{k\, \Delta T}) \T \cdot \M{C}_{z}^{-1} \cdot (\V{d}_{k\, \Delta T} + \M{G}(\V{p})\cdot \V{a}_{k\, \Delta T}) \
\end{equation}

Note that this estimation differs from recent innovations for on-sky
calibration \cite{MeimonFusco2010b, PieralliPuglisi2008a} by the fact
that we cannot choose here the applied commands
$\V{a}_k$. Furthermore, these applied commands $\V{a}_k$ act to
counter the turbulence $w_k$. A good AO correction thus means a
significant correlation between command, $\V{a}_k$, and disturbance,
$\V{z}_k$. Overall, it is important to note that, contrary to the
measurement noise $\V{e}_k$ in a conventional AO, the disturbance
$\V{z}_k$ here is correlated from one subaperture to another (as well
as from one guide star measurement to another guide star measurement
and from one measurement to the next in time) because it includes the
turbulence contribution $\V{w}_k$.

The best parameters $\V{p}^*$ are thus obtained solving
\begin{equation}
  \label{eq:OptimPNoINCR}
  \V{p}^* = {\rm arg ~ min}_{\V{p}} \quad \chi^2_{N, \Delta T} (\V{p})\,,
\end{equation}
taking into account the statistics of $\V{z}_k$ in $\M{C}_{z}$. The
purpose of Sec.~\ref{sec:CovS-Real}, below, is to provide an analysis
of this covariance matrix of the disturbance, in order to understand
the existing correlations in the residuals of this criterion.

\subsection{Covariance expression for $\M{C_{z}}$}
\label{sec:CovS-Real}

From Eq.~(\ref{eq:CovZ}), $\M{C_{z}}$ requires the expression of
$\M{C_{s}}$ and $\M{C_{e}}$. An analytical expression for the
covariance of Shack-Hartmann slopes $\M{C_{s}}$, in square radians,
can be deduced from \cite{Roddier1981a} or \cite{VidalGendron2010a} for
instance. For clarity reasons, we only write in this section the
details of the mathematical derivation for covariance of slopes issued
from the same Shack-Hartmann (same guide star). It can be written
\begin{eqnarray}
  \label{eq:CovxxVidal}
  \Avg{\V{s}_x(i) \cdot \V{s}_x(j) } & = &  \frac {1} {2\,S^2} \, \frac { \partial^2 D_{\phi} } {\partial x^2} (\V{r}_{ij}) \otimes \Pi (\V{r}_{ij}) \otimes \Pi(\V{r}_{ij})  = \mathcal{R}_{Sxx}(\V{r}_{ij}) \\
\label{eq:CovxyVidal}
\Avg{\V{s}_x(i) \cdot \V{s}_y(j) } & = & \frac {1} {2\,S^2} \, \frac { \partial^2 D_{\phi} } {\partial x \, \partial y} (\V{r}_{ij}) \otimes \Pi (\V{r}_{ij}) \otimes \Pi(\V{r}_{ij}) = \mathcal{R}_{Sxy}(\V{r}_{ij}) \\
\label{eq:CovSHyyVidal}
\Avg{\V{s}_y(i) \cdot \V{s}_y(j) } & = & \frac {1} {2\,S^2}  \, \frac { \partial^2 D_{\phi} } {\partial y^2} (\V{r}_{ij}) \otimes \Pi (\V{r}_{ij}) \otimes \Pi(\V{r}_{ij}) = \mathcal{R}_{Syy}(\V{r}_{ij}) \,,
\end{eqnarray}
where $\V{s}_x(i)$ and $\V{s}_y(i)$ are the modeled Shack-Hartmann slopes along $x$ and $y$ directions respectively in the $i$th subaperture (radians),
$\V{r}_{ij}$ is the vector of distance between the centers of $i$th and
$j$th subapertures (meters), $D_{\phi}$ is the structure function of the
phase on the pupil in this guide star direction and $\Pi$ is the
subaperture windowing function, equal to $1$ in the
$[-\sqrt{S}/2;\sqrt{S}/2]\times [-\sqrt{S}/2;\sqrt{S}/2]$ square and equal
to 0 elsewhere, $S$ being the subaperture square area (m$^2$) and $\otimes$
denoting the convolution product. Such model of the Shack-Hartmann slope
measurement takes into account the slope averaging over the subaperture
area, such that $x$-slope in the $i$th subaperture can be written, in
radians, as
\begin{equation}
  \label{eq:SlopeModel}
  \V{s}_x(i) = \frac {1} {S} \, \int \int \frac { \partial \phi} {\partial x} (\V{u}) \, \Pi(\V{u}-\V{r}_i) \, d\V{u}\,, 
\end{equation}
where $\V{r}_i$ is the 2D-coordinate of the center of the $i-$th subaperture.

We recall that the structure function of the wavefront $D_{\phi}$, in
rad$^2$, is
  \begin{equation}
 \label{eq:ScaledSF}
D_{\phi}(r) = \left\{
          \begin{array}{ll}
            6.88 \, (r/r_0)^{5/3} & \mathrm{if}\quad r_0/L_0 =0 \\
            \alpha \, \left(\frac {L_0} {r_0}\right)^{5/3} \left[ 2^{5/6}\, \Gamma\left(\frac{5} {6}\right)-2\, \left(2\, \pi \frac {r} {L_0}\right)^{5/6}K_{5/6}\left(2\, \pi \frac {r} {L_0}\right) \right] ~ & \mathrm{otherwise}\\
          \end{array}
        \right.
\end{equation}
where $K_{5/6}$ is the modified Bessel function of the $3^{\rm rd}$ kind
and of order $5/6$ (Mac Donald's function) and $\alpha = [12/5 \,
\Gamma(6/5)]^{5/6}\,\Gamma(11/6)/\pi^{8/3} \simeq 0.0858$. $L_0$ and $r_0$
are respectively the outer scale and the Fried parameter of the atmospheric
turbulence. The first expression in
Eq.~(\ref{eq:ScaledSF}) stands for Kolmogorov model of turbulence and the
second line stands for von K\'arm\'an model.

Some approximations for the covariance of the Shack-Hartmann slopes
already exist in the literature for Kolmogorov model
\cite{Roddier1981a}.  Another approximation of the slopes variance can
be obtained, for Kolmogorov and von K\'arm\'an turbulence models,
following the Fried model for the Shack-Hartmann and based on phase
differences between the corners of a subaperture.  Such approximation
provides slightly different numerical values than the analysis of
Equations~(\ref{eq:CovxxVidal})-(\ref{eq:CovSHyyVidal}), due to the
approximation of the Fried model neglecting the contribution of higher
frequencies than the cut-off frequency of the lenslet array. This is
why in the following, we prefer to develop our analysis based on
Equations~(\ref{eq:CovxxVidal})-(\ref{eq:CovSHyyVidal}), in order to
study the exact properties of the slopes covariance.

Equations~(\ref{eq:CovxxVidal})-(\ref{eq:CovSHyyVidal}) are not easy
to manipulate and to analyze, but the convolution product being a
product in Fourier domain, the elements of the covariance matrix are
more easily studied below in Fourier domain. The details of the
derivation of the covariance in the Fourier space are gathered in
Appendix~\ref{sec:CovS-Fourier}.


From Fourier-Domain formulation of covariance maps, we can come back
to $(x,y)-$domain and represent the covariance map of slopes for a
Shack-Hartmann. 
\begin{figure}[htb]
  \centering
\begin{minipage}[c]{0.3\linewidth}
\includegraphics[width=5cm, keepaspectratio]{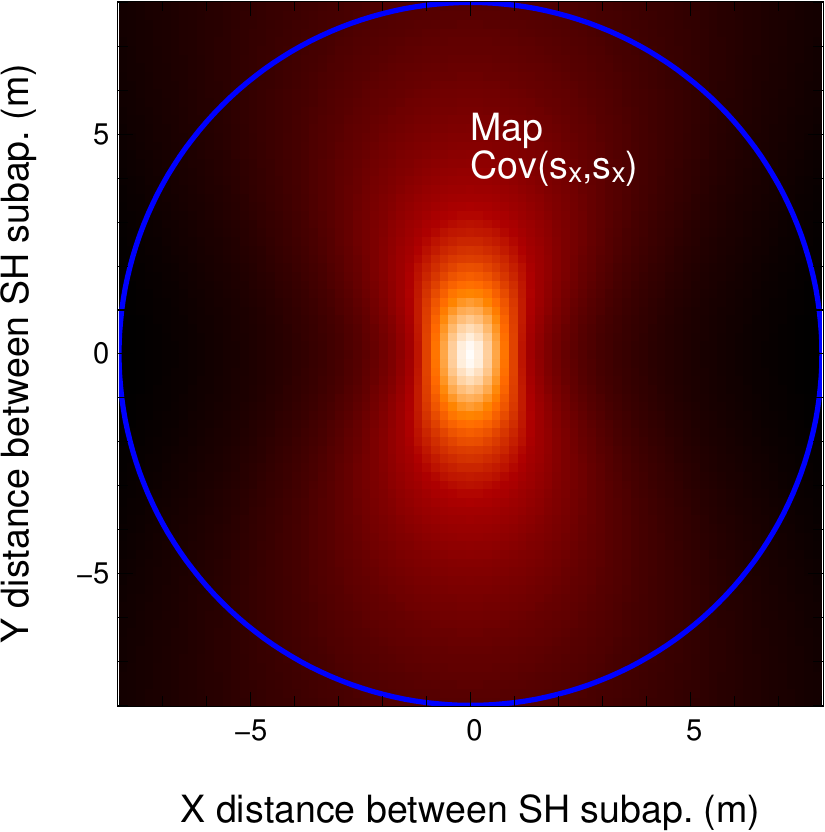}
\end{minipage}
\begin{minipage}{0.3\linewidth}
\includegraphics[width=5cm, keepaspectratio]{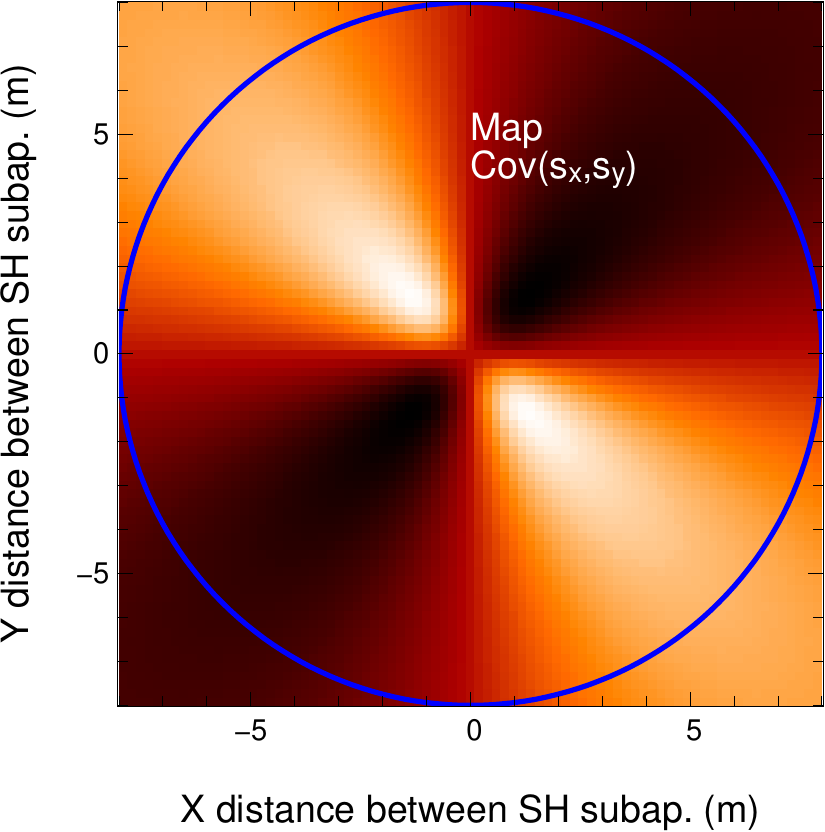}
\end{minipage}  
\begin{minipage}{0.3\linewidth}
\includegraphics[width=5cm, keepaspectratio]{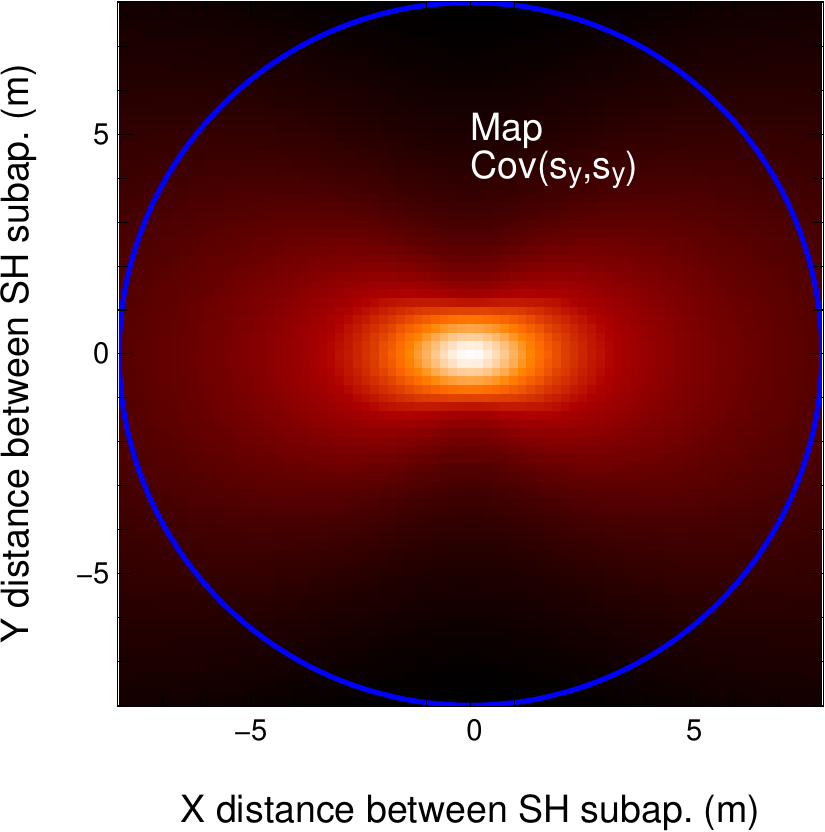}
\end{minipage}  
\caption{\label{fig:CovMapRsXY-AOF}Covariance map of a $40 \times 40$
  shack-Hartmann on an 8m-telescope (like AOF): $\mathcal{R}_{Sxx}$
  (left), $\mathcal{R}_{Sxy}$ (center), and $\mathcal{R}_{Syy}$
  (right). The inside of the circle contains the values that are
  effectively present in a covariance matrix of slopes for a pupil
  diameter of 8~m. }
\end{figure}
Plots of covariance maps of Shack-Hartmann slopes in $(x,y)-$space for
a Shack-Hartmann of subaperture size $\sqrt{S}=0.2$~m over an
8~m-pupil (as in the AOF case) can be numerically obtained, as shown
in Fig.~\ref{fig:CovMapRsXY-AOF}.

\begin{figure}[htb]
  \centering
\begin{minipage}{0.4\linewidth}
\includegraphics[width=7.cm, keepaspectratio]{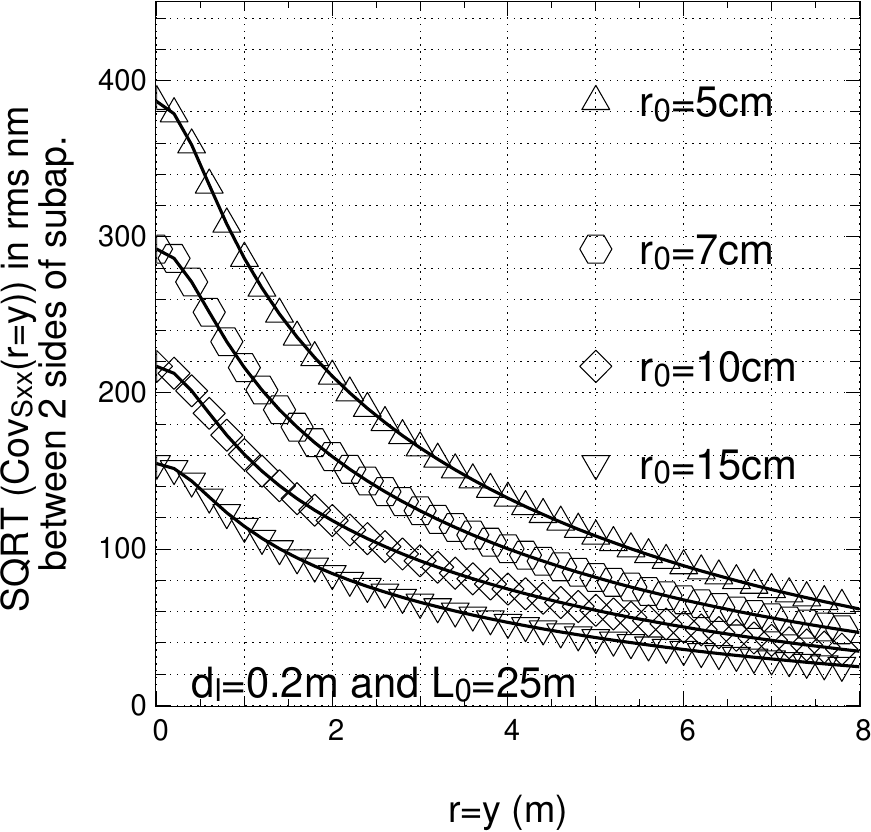}
\end{minipage}
\begin{minipage}{0.4\linewidth}
\includegraphics[width=7.cm, keepaspectratio]{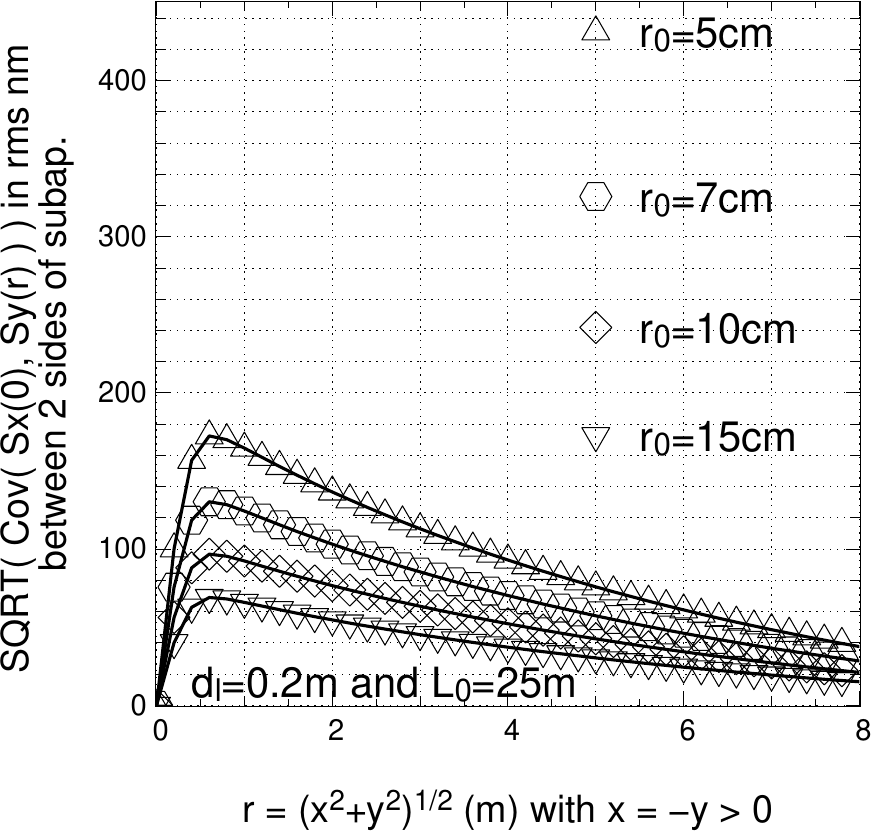}
\end{minipage}  
\caption{\label{fig:CovCut-AOF} Left plot : cut of Cov$(\V{s}_x, \V{s}_x)$
  map in Fig.~\ref{fig:CovMapRsXY-AOF} along $x=0$ and $y \geq 0$. Right
  plot : cut of Cov$(\V{s}_x, \V{s}_y)$ map in
  Fig.~\ref{fig:CovMapRsXY-AOF} along $x = -y \geq 0$. For both plots, the
  square root of the covariance is represented instead of the covariance,
  and it is expressed in nm rms between 2 sides of a subaperture. Various
  colors are used for different $r_0$ rescalings. The markers stand for the
  values effectively appearing in the covariance matrix $\M{C_s}$.}
\end{figure}

From Eqs.~(\ref{eq:SxxRed})-(\ref{eq:SyyRed}) in
Appendix~\ref{sec:CovS-Fourier}, and the linearity of the Fourier
transform, we can see that $\mathcal{R}_{Sxx}$, $\mathcal{R}_{Sxy}$
and $\mathcal{R}_{Syy}$ are proportional to $r_0^{-5/3}$ via the
turbulence spectrum $\mathcal{S}_{\phi}$. In addition, the dependence
on $L_0$ is only visible at low frequencies ($<\, 1/L_0$).  In order
to understand how much the slopes are correlated over the pupil, for
every guide star, we represent in Fig.~\ref{fig:CovCut-AOF} cuts of the
left and center maps of Fig.~\ref{fig:CovMapRsXY-AOF}, where they
exhibit the highest correlation. For the left map of
Fig.~\ref{fig:CovMapRsXY-AOF}, the cut is made for $x=0$ and $y \geq
0$ and presented on the left plot of Fig.~\ref{fig:CovCut-AOF} for
various $r_0$ values. For central map of
Fig.~\ref{fig:CovMapRsXY-AOF}, the cut is made for $x = -y \geq 0 $
and is presented on the right plot of Fig.~\ref{fig:CovCut-AOF} for
the same $r_0$ values.

The values observed in Fig.~\ref{fig:CovCut-AOF} allow to estimate the
order of magnitude of the $\M{C}_s$ contribution to $\M{C}_z$ matrix,
compared to the measurement noise part $\M{C}_e$. In the context of the
AOF, with the laser guide stars and the wavefront sensors design, the
measurement noise has been estimated in \cite{BechetTallon2012} to be of the
order of 50 to 150~nm rms between two subaperture sides. This means that
$\M{C}_s$ contribution to the total variance of $\M{C}_z$ in more than 50\%
rms even in the optimistic case of an $r_0$ value of 16~cm. It also shows
that the covariance matrix $\M{C}_z$ could not be reasonably approximated by a
diagonal, because this contribution from $\M{C}_s$ to the non diagonal
terms of $\M{C}_z$ is large in comparison with the variance value. As a
matter of fact, the covariance presented in the left plot
of Fig.~\ref{fig:CovCut-AOF} decreases very slowly with the distance $r$
between 2 subapertures. The covariance is equal to the variance divided by
2 only for $r$ beyond 1.2~m (6 subapertures), and divided by 4 for $r$
beyond 2~m (10 subapertures).

As a conclusion, the covariance matrix $\M{C}_z$ has a complex
structure, and is far from being diagonal. A good estimator based on
the solving of Eq.~(\ref{eq:OptimPNoINCR}) would need to take these
noise correlations into account. All this leads us to derive another
measurement model in Section~\ref{sec:DiffMeasEq}, in order to
overcome these difficulties.

\section{Criterion from increments of the measurement equation}
\label{sec:DiffMeasEq}

Instead of using Eq~(\ref{eq:MeasEqZ}), we use now the difference
between two successive measurements,
\begin{equation}
  \label{eq:diffMeasEqZ}
  \delta \V{d}_k = \V{d}_{k+1} - \V{d}_k = -\M{G}\cdot \delta\V{a}_k + \delta\V{z}_k
\end{equation}
where $\delta\V{a}_k=\V{a}_{k+1}-\V{a}_k$ and
$\delta\V{z}_k=\V{z}_{k+1}-\V{z}_k$.  

The idea of such approach is to transform the estimation problem
keeping its linearity and Gaussian statistics properties, with a
disturbance covariance matrix $\M{C}_{\delta z}$ which could be better
approximated by a diagonal matrix and decorrelating (approximately)
signal ($\M{G}\cdot \delta\V{a}_k$) and disturbance ($\delta\V{z}_k$).

In the same way as in Sec.~\ref{sec:MeasEq}, we justify this approach with
an analysis of the covariance matrix of the disturbance, here
$\M{C}_{\delta z}$. The disturbance $\delta\V{z}_k$ follows a zero-mean
Gaussian statistics of covariance matrix
\begin{eqnarray}
  \label{eq:CovdeltaZ}
  \M{C}_{\delta z} & = & \Avg{ (\V{z}_{k+1} - \V{z}_k)  \cdot (\V{z}_{k+1} - \V{z}_k) \T } = \Avg{\M{S}(\delta w_k) \cdot {\M{S}(\delta w_k)}\T} + \Avg{\delta \V{e}_k \cdot \delta \V{e}_k\T} \\
\label{eq:CovdeltaZDev2Ce}
~ & = & \M{C}_{\delta s} + 2 \, \M{C}_e\,.
\end{eqnarray}

\subsection{Criterion}
\label{sec:CritDiffMeasEq}

Using only one set of incremental measurements $\delta \V{d}_k$, one can
fit the interaction matrix parameters $\V{p}$ using maximum likelihood approach,
\ie minimizing
\begin{equation}
  \label{eq:Chi2CBE}
  \chi_{1}^2 (\V{p}) = (\delta \V{d}_k + \M{G}(\V{p})\cdot \delta \V{a}_k) \T \cdot \M{C}_{\delta z}^{-1} \cdot (\delta \V{d}_k + \M{G}(\V{p})\cdot \delta \V{a}_k) \,,
\end{equation}
where the lower indice $1$ of the $\chi_{1}^2$ stands for using only
one set of incremental measurements and commands. Selecting several
sets of incremental measurements separated by a certain number of
frames, $\Delta T$, large enough to avoid correlations between the
associated $\delta \V{z}_k$ and $\delta \V{z}_{k+\Delta T}$, more
residuals can be added in the $\chi^2$ computation, and the criterion
can be changed to
\begin{equation}
  \label{eq:Chi2CBE-N}
  \chi_{N,\Delta T}^2 (\V{p}) = \sum_{k=1}^{k=N}(\delta \V{d}_{k\, \Delta T} + \M{G}(\V{p})\cdot \delta \V{a}_{k\, \Delta T}) \T \cdot \M{C}_{\delta z}^{-1} \cdot (\delta \V{d}_{k\, \Delta T} + \M{G}(\V{p})\cdot \delta \V{a}_{k\, \Delta T}) \
\end{equation}

In other words, the method now estimates that the best interaction
matrix parameters are
\begin{equation}
  \label{eq:OptimPCBE}
  \V{p}^* = {\rm arg ~ min}_{\V{p}} \quad \chi^2_{N, \Delta T} (\V{p})\,.
\end{equation}
Equations~(\ref{eq:Chi2CBE-N}) and (\ref{eq:OptimPCBE}) characterize
the identification method successfully studied in simulations
\cite{BechetTallon2012}. In order for this method to be efficient, we
need to :
\begin{enumerate}
\item Know, in a good approximation, $\M{C}_{\delta z}$. Here approximated
  by $2\, \M{C}_e$.
\item Demonstrate that  $\V{\delta a}_k$ is not correlated to $\V{\delta z}_k$.
\end{enumerate} 

These 2 points are critical to be able to claim with how much
precision we could estimate the system parameters. The theoretical
analysis below clarifies the first one of these two points, with the result
of Eq.~(\ref{eq:ApproxCdz}). The second point is still not clearly analyzed
from the theory.

\subsection{Covariance of incremental Shack-Hartmann slopes}
\label{sec:DiffCovMap}

\begin{figure}[h]
  \centering
\begin{minipage}[c]{0.3\linewidth}
\includegraphics[width=5cm, keepaspectratio]{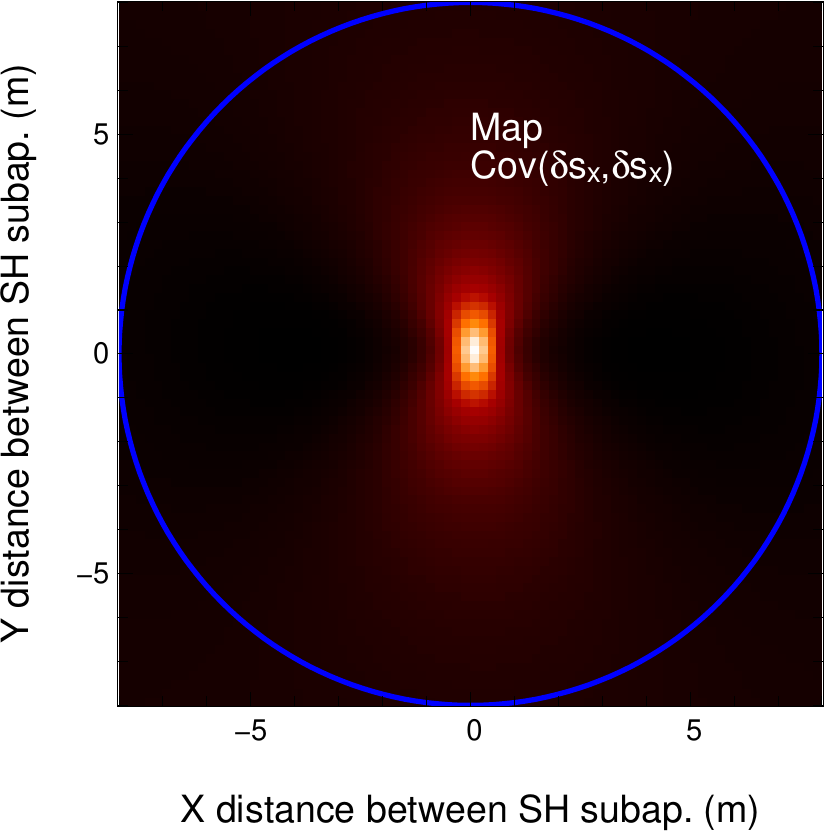}
\end{minipage}
\begin{minipage}{0.3\linewidth}
\includegraphics[width=5cm, keepaspectratio]{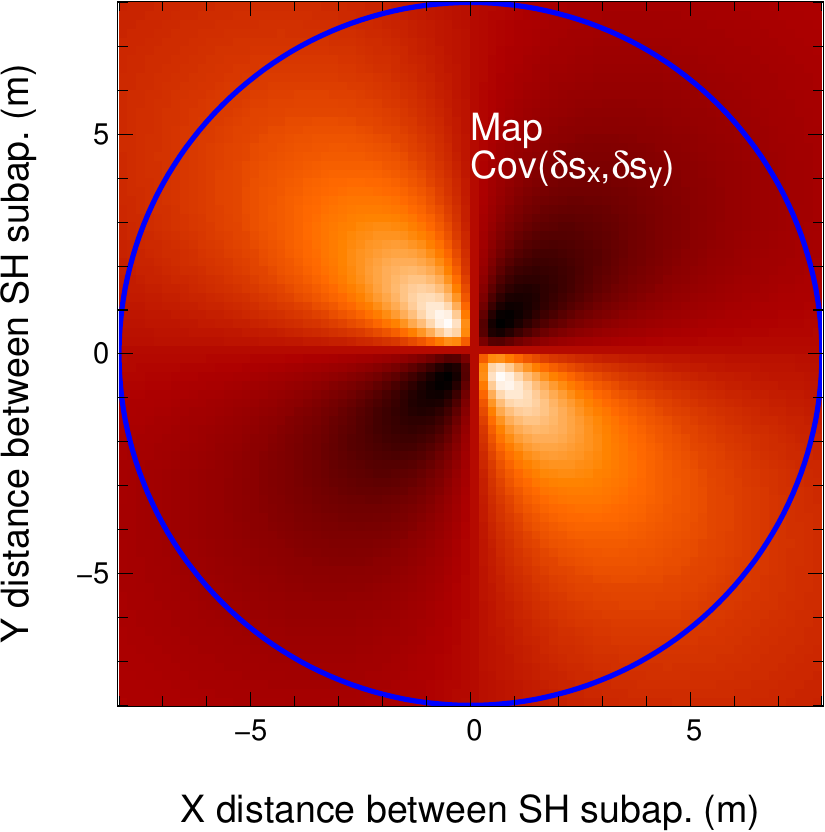}
\end{minipage}  
\begin{minipage}{0.3\linewidth}
\includegraphics[width=5cm, keepaspectratio]{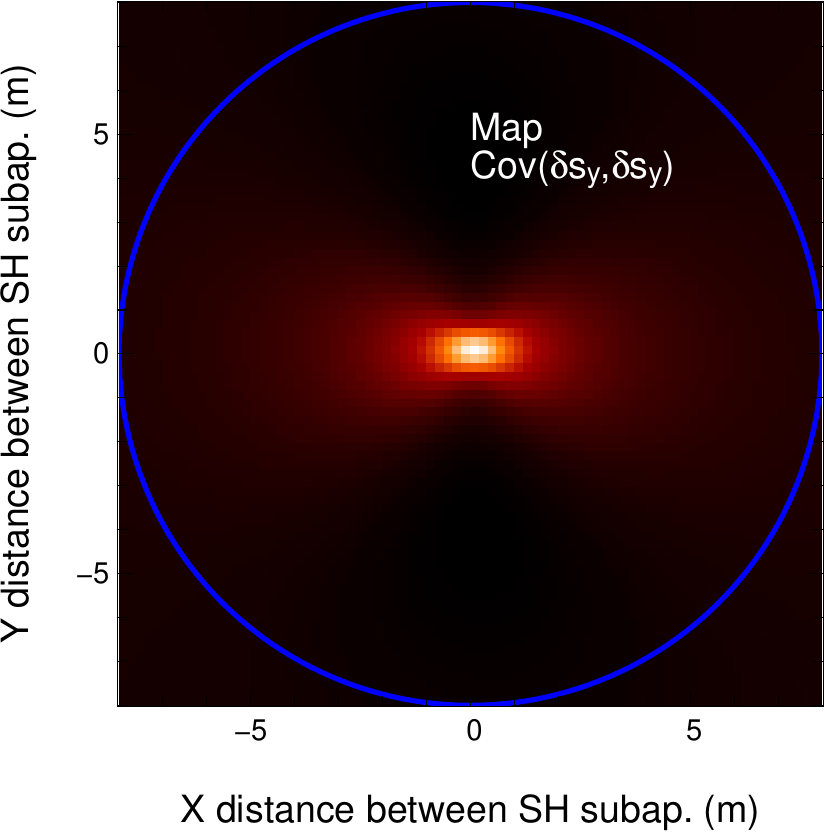}
\end{minipage}  
\caption{\label{fig:Rds-vaof} Covariance maps in $(x,y)-$space
  $\M{R}_{\delta Sxx}$, $\M{R}_{\delta Sxy}$ and $\M{R}_{\delta Syy}$ of
  the incremental slopes are represented in AOF-like conditions. }
\end{figure}

As for the covariance map of th Shack-Hartmann slopes in
Sec.~\ref{sec:CovS-Real}, Fourier space analysis is again used to model
$\M{C}_{\delta z}$. The details of the mathematical derivation are
gathered in Appendix~\ref{sec:DiffFourierCovMap}. Once the
Fourier-domain spectral density maps of the incremental slopes are
computed, we can come back to $(x,y)-$space, to get the expected
covariance maps over the aperture. We obtain maps like in
Figs.~\ref{fig:Rds-vaof} estimated for typical configuration of the
AOF system at the VLT.

\begin{figure}[h]
  \centering
\begin{minipage}[c]{0.3\linewidth}
\includegraphics[width=5.3cm, keepaspectratio]{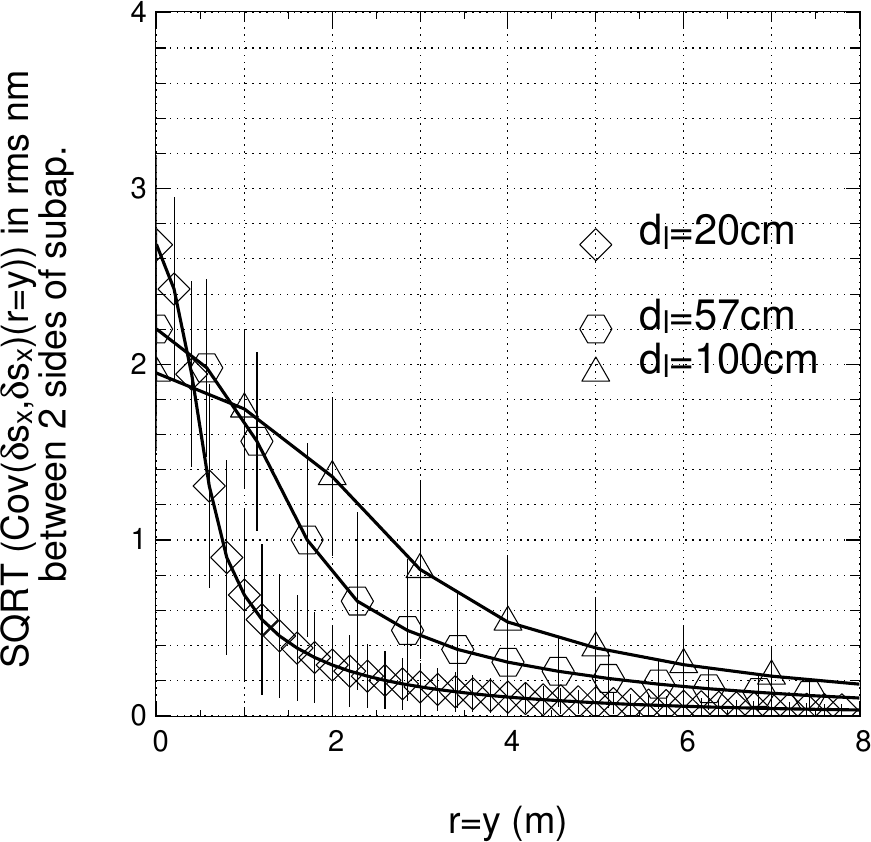}
\end{minipage}
\begin{minipage}{0.3\linewidth}
\includegraphics[width=5.3cm, keepaspectratio]{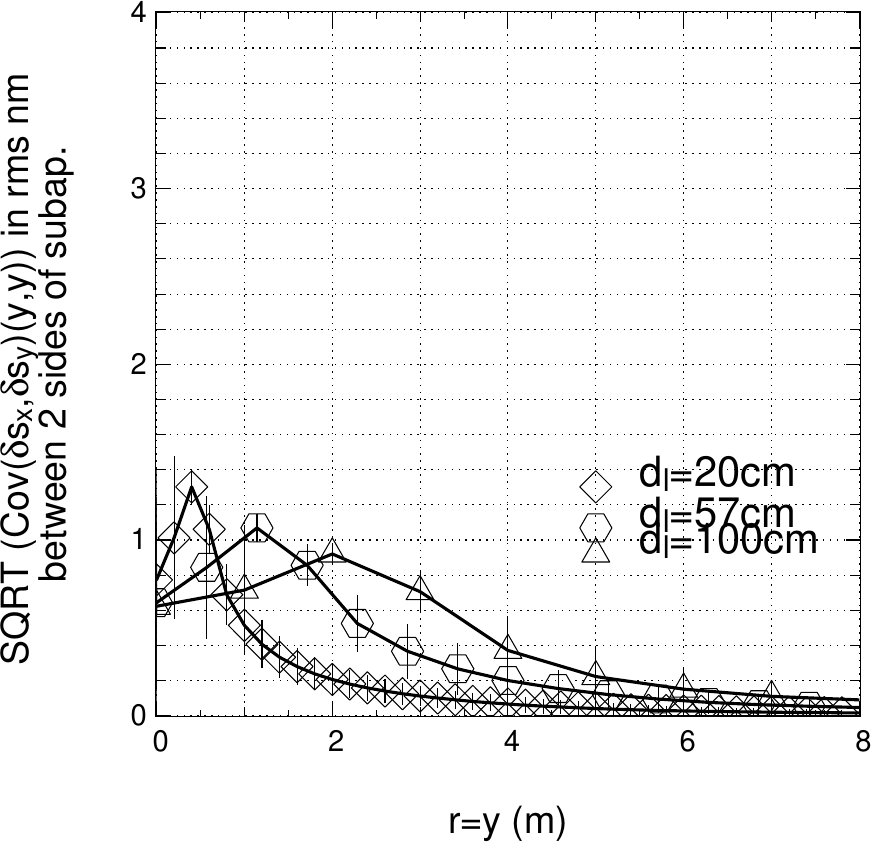}
\end{minipage}  
\begin{minipage}{0.3\linewidth}
\includegraphics[width=5.3cm, keepaspectratio]{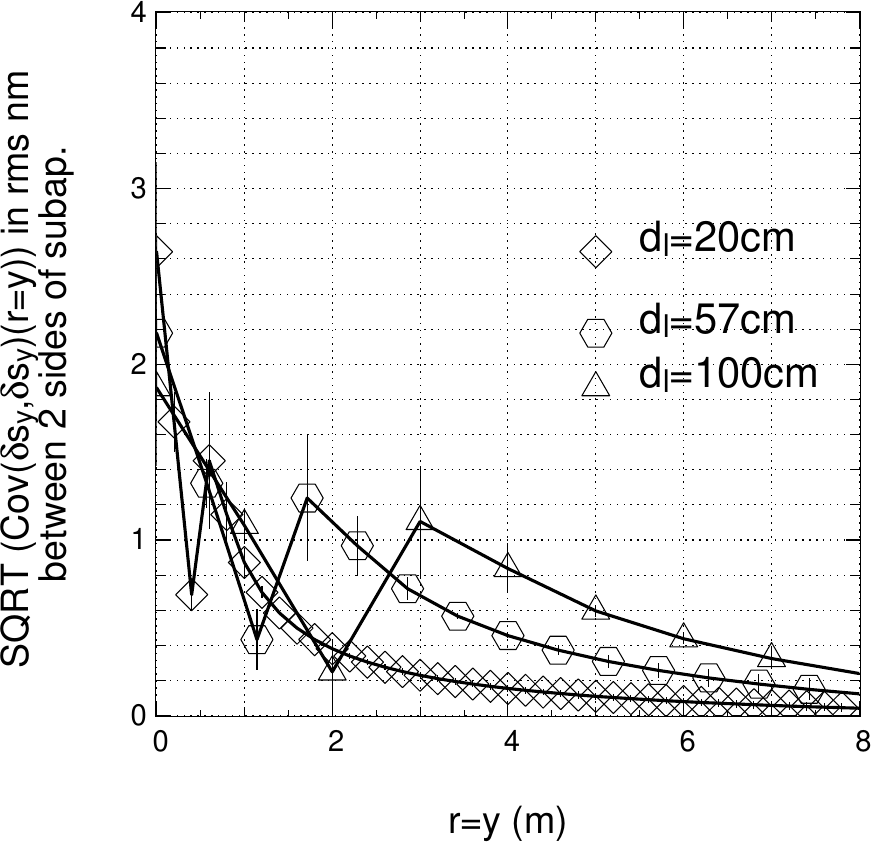}
\end{minipage}  
\caption{\label{fig:Rds-vaof-cuts-d} Cut of covariance maps like the
  one in Fig.~\ref{fig:Rds-vaof}, along $y>0$ and $x=0$ axis for left
  and right plot, and along $x=y\geq 0$ axis for the plot in the
  middle. The square root of the Covariance is plotted, in order to
  express it in nm rms, and to ease the comparison with noise
  level. WFS exposure time is 1~ms. The curves with different markers
  stand for different subaperture sizes $d_l$: 1~m (triangles), 57~cm
  (hexagons) and 20~cm (diamonds).}
\end{figure}

\begin{figure}[h]
  \centering
\begin{minipage}[c]{0.3\linewidth}
\includegraphics[width=5.3cm, keepaspectratio]{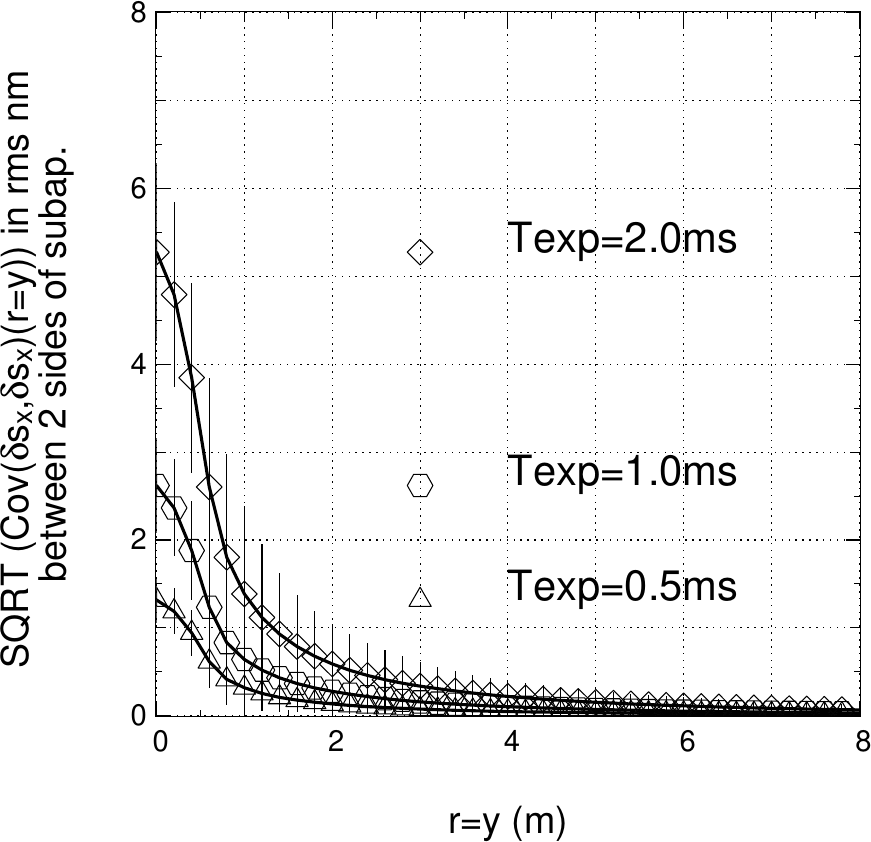}
\end{minipage}
\begin{minipage}{0.3\linewidth}
\includegraphics[width=5.3cm, keepaspectratio]{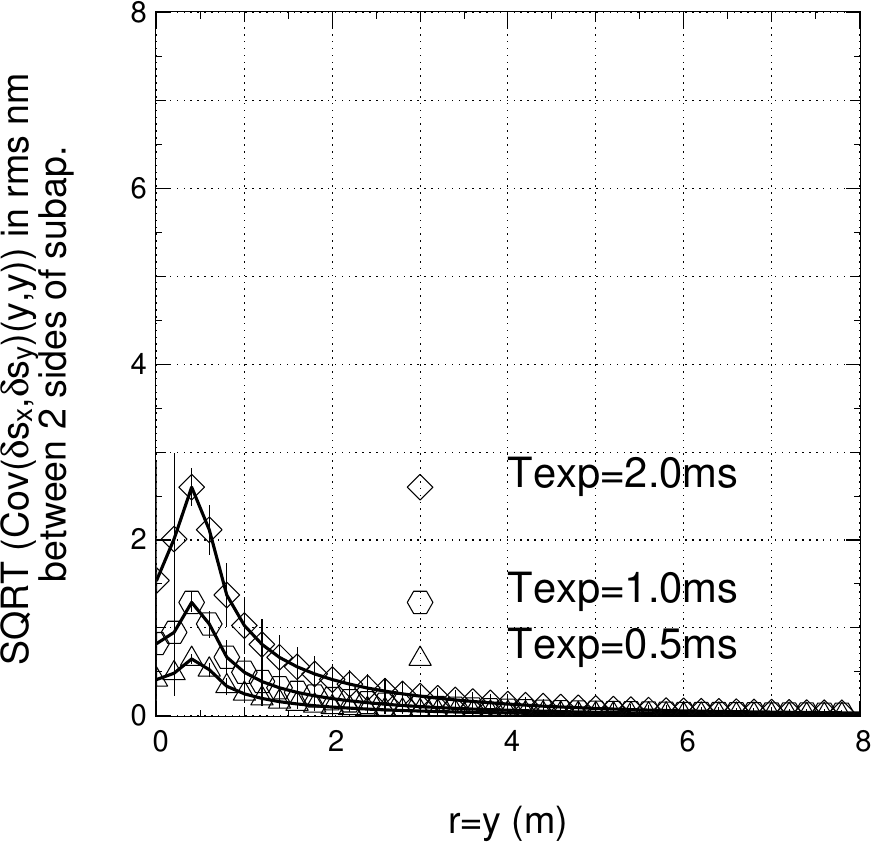}
\end{minipage}  
\begin{minipage}{0.3\linewidth}
\includegraphics[width=5.3cm, keepaspectratio]{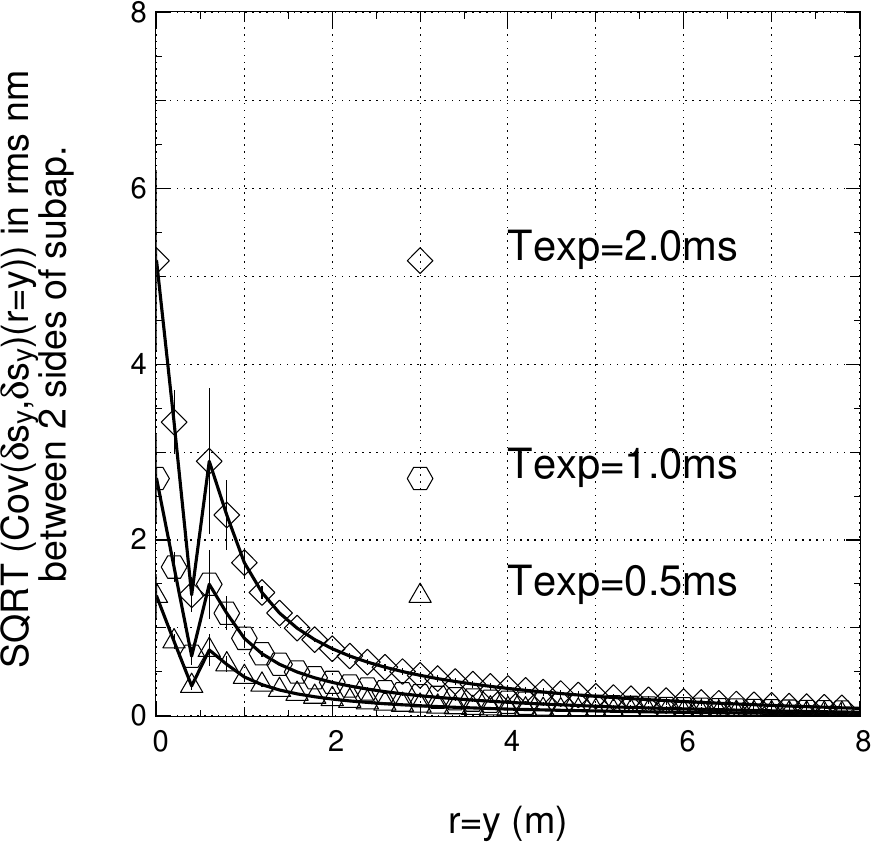}
\end{minipage}  
\caption{\label{fig:Rds-vaof-cuts-Texp} Cut of AOF-like covariance
  maps from Fig.~\ref{fig:Rds-vaof}, along $y>0$ and $x=0$ axis for
  left and right plot, and along $x=y\geq 0$ axis for the plot in the
  middle. The square root of the Covariance is plotted, in order to
  express it in nm rms, and to ease the comparison with noise
  level. The curves with different markers stand for different WFS
  exposure times: 0.5~ms (triangles), 1~ms (hexagons) and 2~ms
  (diamonds).}
\end{figure}

This contains exactly the values that are present in the covariance
matrix $\M{C}_{\delta s}$ we are searching for. Some $x-$ and $y-$cuts
are done over these maps, and are represented in
Figs.~\ref{fig:Rds-vaof-cuts-d} and
\ref{fig:Rds-vaof-cuts-Texp}. Abscissas are in meters, and ordinates
in square nanometers between two side of a subaperture. It has been
observed that, when $r_0$, $L_0$ and $\tau_0$ change, the shape of
these cuts stay similar, and the level does not change much, staying
in the range of 5 to 50~nm$^2$. In particular,
Fig.~\ref{fig:Rds-vaof-cuts-d} enhance that the reduction of the
subaperture size from 57~cm, like for NAOS, to 20~cm, like for the
AOF, significantly accelerates the decorrelation of $\V{\delta z}$
when considering farther subapertures. From
Fig.~\ref{fig:Rds-vaof-cuts-Texp}, one can observe the increase of the
correlation with the WFS exposure time. However, in any case from
0.5~ms to 2ms of exposure, the covariance quickly drops down after 1
meter of distance.

More time would be required to really get a quantitative evaluation of
these covariance coefficients as a function of the atmospheric
parameters. Nevertheless, one observes that beyond 1 meter of distance
the covariance is below a few nm$^2$, and even at its maximum values
it is below the level of expected measurement noise in the AOF system
($\sim$80~nm rms).

These examples of covariance values for the incremental slopes of a
Shack-Hartmann provide order of magnitude for the diagonal elements
and off-diagonal elements of the covariance matrix $\M{C}_{\delta
  s}$. It confirms that the contribution of the off-diagonal elements
decreases faster than for $\M{C}_{s}$ with respect to the maximum
diagonal ones, when the distance between the subapertures
increases. This can be clearly seen in Figs.\ref{fig:Rds-vaof-cuts-d},
where the decorrelation appears even steeper for subaperture size of
$d = 0.2~$m (AOF-like), than for subaperture size of $d = 0.57~$m
(NAOS-like). In addition, the amplitude of the diagonal elements of
the covariance matrix is also small compared to the expected
measurement noise level we mentioned for the AOF system. 

\begin{figure}[h]
  \centering
\begin{minipage}[c]{0.3\linewidth}
\includegraphics[width=5cm, keepaspectratio]{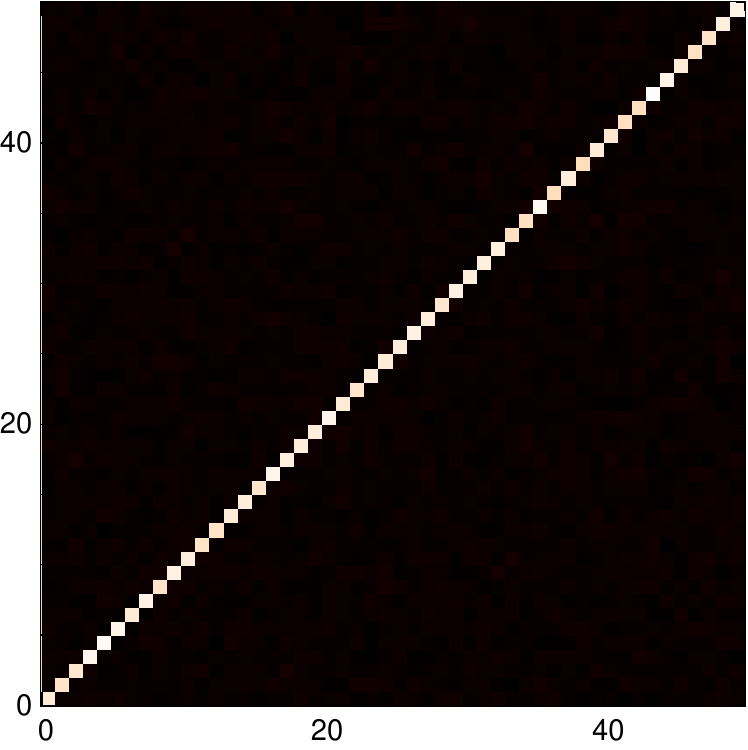}
\end{minipage}
\begin{minipage}{0.3\linewidth}
\includegraphics[width=5cm, keepaspectratio]{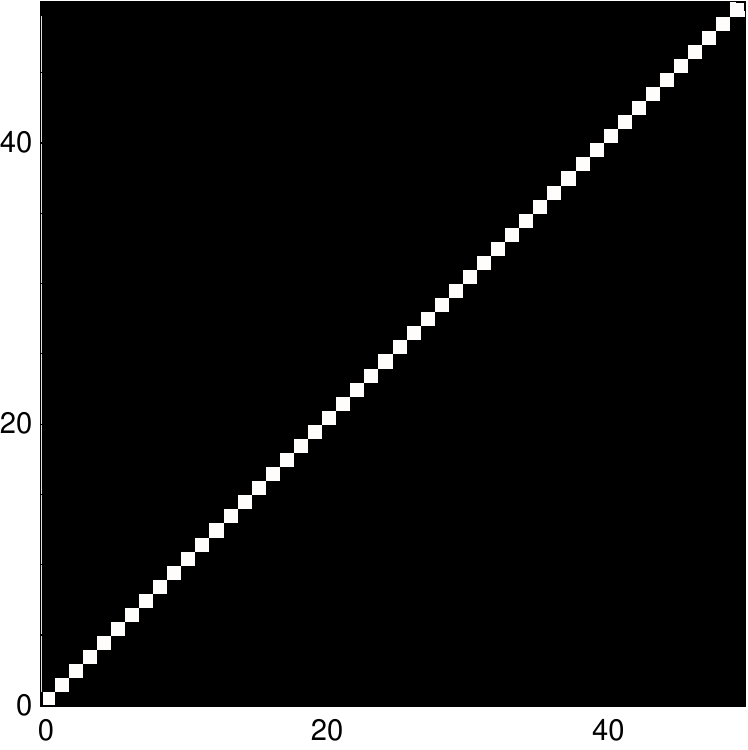}
\end{minipage}  
\begin{minipage}{0.3\linewidth}
\includegraphics[width=5cm, keepaspectratio]{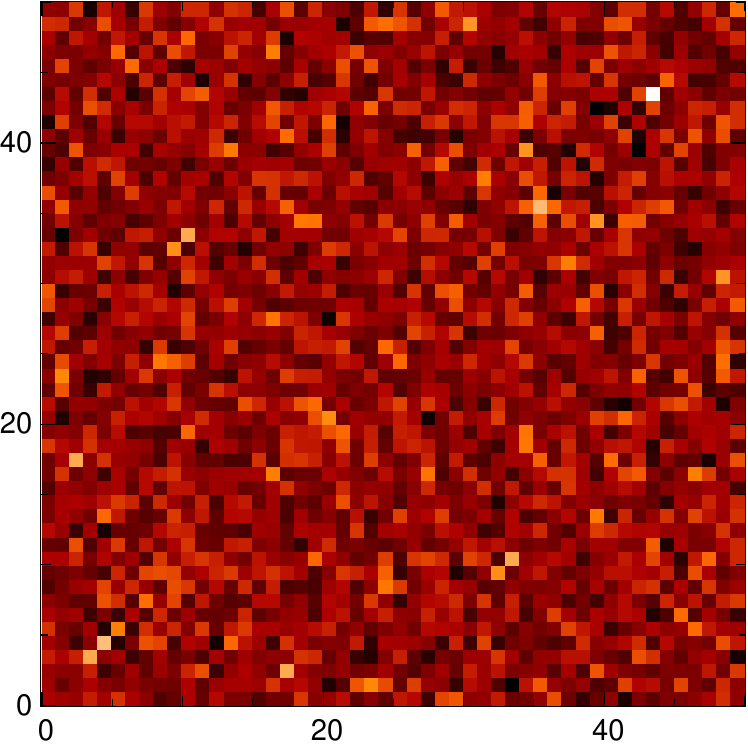}
\end{minipage}  
\caption{\label{fig:ExCovdz}Example of estimation of covariance matrices
  deduced from an open-loop sequence of Shack-Hartmann measurements of an
  evolving turbulent atmosphere (only the 50 first rows and columns of the
  full $2304 \times 2304$ matrices on the $40 \times 40$ subapertures SH
  are displayed): $\M{C}_{\delta z}$ (left), $\M{C}_e$ (middle) and
  $\M{C}_{\delta s}$ (right). The non zero (diagonal) values of $\M{C}_e$
  are $\sigma^2_e=6800~$mas$^2$ per subaperture (equivalent to $\sim$80~nm
  rms between 2 subaperture sides), while the noisy shape of $\M{C}_{\delta
    s}$ has values of the order of 5\% of this, \ie between $-300$ and
  $+300$mas$^2$.The contribution of the right matrix to the covariance
  $\M{C}_{\delta z}$ on the left is thus quite small. }
\end{figure}

In the case of incremental slopes, we can therefore approximate the
covariance matrix $\M{C}_{\delta z}$ by twice the diagonal covariance
of the measurement noise $\M{C}_e$, \textit{i.e.}
\begin{equation}
  \label{eq:ApproxCdz}
  \M{C}_{\delta z} \simeq 2\, \M{C}_e\,.
\end{equation}
Eq.~(\ref{eq:CovdeltaZDev2Ce}) is then replaced by the approximation
in Eq.~(\ref{eq:ApproxCdz}). As another example, Fig.\ref{fig:ExCovdz}
presents the three matrices $\M{C}_{\delta z}$, $\M{C}_{\delta s}$ and
$\M{C}_{e}$ as numerically estimated from the simulation of a
closed-loop AO system. From this simulation, we observe that the non
diagonal terms of $\M{C}_{\delta z}$ are about 20 times lower than the
diagonal ones. The noise measurement dominates in $\delta \V{z}$.

\section{Implementations}
\label{sec:Implem}

Another theoretical analysis of this identification procedure has been
done, which concerns the comparison of different implementations of
the identification method introduced in
Sec.~\ref{sec:DiffMeasEq}. Various implementations have been used for
the AOF study and the aim of this section is to clarify what can be
said about these different implementations and the reasons why they
can provide identical or different results.

\subsection{Direct computation of misregistration parameters}
\label{sec:CBE}

This implementation consists in a model-fitting directly defined by
Equations~(\ref{eq:Chi2CBE-N}) and (\ref{eq:OptimPCBE}). It is also
the implementation used in the simulations presented in B\'echet
\textit{et al.} \cite{BechetTallon2012}. The interaction matrix is
defined as a parametric model $\M{G}(\V{p})$, build from what we know
about DM influence functions and shack-Hartmann linearized
modeling. The criterion of Eq.~(\ref{eq:Chi2CBE-N}) is directly
minimized with respect to the parameters $\V{p}$, using a non-linear
optimization algorithm, which is a Levenberg-Marquardt algorithm
modified to take into account a \textit{trust region}.

\subsection{Computation of misregistration parameters after an
  interaction matrix estimation}
\label{sec:JKO}

The criterion mentioned in Eq.~(\ref{eq:Chi2CBE-N}) is general, so that it
could be applied to some ten parameters like misalignments characteristics
for instance, as well as to the full interaction matrix coefficients such
that each element of $\V{p}$ is a coefficient $\M{G}_{i,j}$.

We could then rewrite the criterion of Eq.~(\ref{eq:Chi2CBE-N})
\begin{equation}
  \label{eq:Chi2CBE-N-G}
  \chi_{N,\Delta T}^2 (\M{G}) =  \frac{1} {2} \, \sum_{k=1}^{k=N}(\delta \V{d}_{k\, \Delta T} + \M{G}\cdot \delta \V{a}_{k\, \Delta T}) \T \cdot \M{C}_{\delta z}^{-1} \cdot (\delta \V{d}_{k\, \Delta T} + \M{G}\cdot \delta \V{a}_{k\, \Delta T}) \  
\end{equation}

In this last case, the criterion~(\ref{eq:Chi2CBE-N-G}) becomes a quadratic
function of $\M{G}$, for which the solution 
\begin{equation}
  \label{eq:OptimGCBE}
  \M{G}^* = {\rm arg ~ min}_{\M{G}} \quad \chi^2_{N, \Delta T} (\M{G})\,
\end{equation}
is obtained when the first derivative of the criterion is zero, \ie
\begin{equation}
  \label{eq:1rstDeriv0}
  \frac {\partial \chi_{N,\Delta T}^2} {\partial \M{G}} = \M{C}_{\delta z}^{-1} \, \sum_{k=1}^{k=N} (\delta \V{d}_{k\, \Delta T} + \M{G}\cdot \delta \V{a}_{k\, \Delta T})  \cdot  \delta \V{a}_{k\, \Delta T}\T = 0\,.
\end{equation}


Since $\M{C}_{\delta z}$ is a positive definite covariance matrix
(symmetric), its inverse is also symmetric positive definite, and then the
solution $\M{G}^*$ verifies
\begin{equation}
  \label{eq:GsolCBE}
  \M{G}^* \cdot \sum_{k=1}^{k=N} (\delta \V{a}_{k\, \Delta T}  \cdot \delta \V{a}_{k\, \Delta T}\T)  = - \sum_{k=1}^{k=N} (\delta \V{d}_{k\, \Delta T} \cdot \delta \V{a}_{k\, \Delta T}\T )\,. 
\end{equation}

We choose the following notations for approximated estimates of the
covariance matrices
\begin{equation}
  \label{eq:Covaa-Covda}
  \M{C}^*_{\delta a, \delta a} = \frac {1} {N} \, \sum_{k=1}^{N} (\delta \V{a}_{k\, \Delta T}  \cdot \delta \V{a}_{k\, \Delta T}\T) \quad {\textrm{and}} \quad
  \M{C}^*_{\delta d, \delta a} = \frac {1} {N} \, \sum_{k=1}^{N} (\delta \V{d}_{k\, \Delta T} \cdot \delta \V{a}_{k\, \Delta T}\T )\,.
\end{equation}
In case $\M{C}^*_{\delta a, \delta a}$ is invertible, then the
solution $\M{G}^*$ is
\begin{equation}
  \label{eq:solGinvert}
  \M{G}^* = - \M{C}^*_{\delta d, \delta a}\, \cdot \M{C}^{*\, -1}_{\delta a, \delta a} \, .
\end{equation}
In general, it is difficult to ensure that $\M{C}^*_{\delta a, \delta
  a}$ will be invertible (\textit{e.g.} some actuators may not be
commanded or the recorded sequence is not long enough). As a
consequence, the inverse above will be replaced by the generalized
inverse of $\M{C}^*_{\delta a, \delta a}$, computed by Truncated
Singular Value Decomposition (T-SVD). In such case, it is
mathematically equivalent to compute the matrix estimate $\M{G}^*$,
following the procedure below. This procedure, suggested by J.~Kolb
\cite{KolbEtAl2012}, computes the matrix $\M{G}^{*}$ using
the expression
\begin{equation}
  \label{eq:GJKO}
  \M{G}^{*} = - \Delta_d \cdot {\Delta_a}^{\dag}\, 
\end{equation}
where the $\Delta_d$ and $\Delta_a$ matrices consist of concatenation of incremental vectors $\delta \V{d}_k$ of measurements and $\delta \V{a}_k$ of commands
respectively, in $N$ columns, as illustrated below
\begin{eqnarray}
  \label{eq:DeltaDN}
  \Delta_d & = & \left[ \delta \V{d}_{\Delta T} ~|~ \delta \V{d}_{2 \Delta T} ~|~ \delta \V{d}_{3 \Delta T} ~|~ ...~ |~ \delta \V{d}_{N \Delta T} \right]\\
\quad & ~ & \nonumber \\
  \label{eq:DeltaAN}
  \Delta_a & = & \left[ \delta \V{a}_{\Delta T} ~| ~\delta \V{a}_{2 \Delta T}~ |~ \delta \V{a}_{3 \Delta T} ~|~ ...~ |~ \delta \V{a}_{N \Delta T} \right]\,,
\end{eqnarray}
and ${\Delta_a}^{\dag}$ is a general inverse of $\Delta_a$ obtained by
Truncated Singular Value Decomposition (T-SVD).

The mathematical equivalence between Eqs.~(\ref{eq:solGinvert}) and
(\ref{eq:GJKO}) is proved in Appendix\ref{sec:equivKolb}. In addition,
we have checked with our simulations that the two computational
approaches provide the same estimate for the interaction matrix
$\M{G}^{*}$.

\subsection{Precision on this intermediate interaction matrix estimate}
\label{sec:precisionG}

The accuracy of the estimation of the coefficients of the interaction
matrix $\M{G}^{*}$, by the method described above, can be
theoretically formulated thanks to the Hessian of the criterion. The
coefficients of the inverse of the Hessian define the error bars
(diagonal coefficients) and the correlations (off-diagonal
coefficients).

More work is required to derive an exact expression for this precision, but
in a first approximation it would lead to show that the error bar on
coefficient $\M{G}_{i,j}$ is $1/\sigma^2_{\delta aj}$, the inverse of the
variance of the incremental commands on the $j-$th actuator.

In other words, the error bars are the same in a given column of the
interaction matrix, but it can be different from one column to
another. Another remark is that the higher the variance of incremental
commands, the better the precision. This explains why when the system tends
to become unstable, the estimation of the interaction matrix on sky in
closed-loop could be more precise, thanks to some particular highly excited
actuators.

\section{Conclusion}
\label{sec:Conclu}

This paper presents the recent advances obtained in the analysis of
the identification method for misregistration parameters in
AO. Following the promising results previously obtained on simplified
simulations \cite{BechetTallon2012}, a theoretical analysis of the
estimation method, its assumptions and its accuracy needed to be
addressed. 

Thanks to both analytical and numerical analysis, the paper enhances
the benefit of the method based on the \textit{incremental} measurement
equation, compared to the one based on the classical measurement
equation. Covariances and correlations between signal and disturbance
in this case appear easier to compute and to approximate, in order to
ensure the use of a reliable estimator. 

Two implementations of this identification method are also discussed,
and the equivalence between two algorithms to estimate an intermediate
estimation of the interaction matrix is proved. This is important
since the identification of misregistration parameters requires to be
applied in real time, and thus computational aspects must be
considered to select an efficient algorithm even for large AO
systems.

Finally, the report show perspectives to derive a theoretical analysis
of the accuracy through the study of the Hessian matrix. However, more
theoretical work is required to reach such a goal.

\appendix    


\section{Fourier-domain covariance of Shack-Hartmann slopes}
\label{sec:CovS-Fourier}

We express the Shack-Hartmann slopes covariances of
Eqs.~(\ref{eq:CovxxVidal}-\ref{eq:CovSHyyVidal}) using their power spectral
density in the Fourier domain
\begin{eqnarray}
  \label{eq:SxxVidal}
  \mathcal{S}_{Sxx}(\V{\kappa}) = \mathcal{F}(\mathcal{R}_{Sxx}) & = &  \frac {1} {2\, S^2} \, \mathcal{F}\left(\frac { \partial^2 D_{\phi} } {\partial x^2}\right) \, \left[\mathcal{F}\left(\Pi \right)\right]^2 \\
\label{eq:SxyVidal}
\mathcal{S}_{Sxy}(\V{\kappa}) = \mathcal{F}(\mathcal{R}_{Sxy}) & = & \frac {1} {2\, S^2} \,  \mathcal{F}\left(\frac { \partial^2 D_{\phi} } {\partial x \, \partial y} \right) \, \left[\mathcal{F}\left(\Pi \right)\right]^2  \\
\label{eq:SyyVidal}
\mathcal{S}_{Syy}(\V{\kappa}) = \mathcal{F}(\mathcal{R}_{Syy}) & = & \frac {1} {2\, S^2} \, \mathcal{F}\left(\frac { \partial^2 D_{\phi} } {\partial y^2} \right)\, \left[\mathcal{F}\left(\Pi \right)\right]^2 \,.
\end{eqnarray}

We can use the properties of the Fourier Transform to write
\begin{eqnarray}
  \label{eq:FourierProperties1}
\mathcal{F}\left(\frac { \partial^2 D_{\phi} } {\partial x^2}\right) & = & -4\pi^2 \kappa_x^2 \mathcal{F}(D_{\phi}) \\
  \label{eq:FourierProperties2}
\mathcal{F}\left(\frac { \partial^2 D_{\phi} } {\partial x \partial y}\right) & = & -4\pi^2 \kappa_x \kappa_y \mathcal{F}(D_{\phi}) \\
  \label{eq:FourierProperties3}
\mathcal{F}\left(\frac { \partial^2 D_{\phi} } {\partial y^2}\right) & = & -4\pi^2 \kappa_y^2 \mathcal{F}(D_{\phi}) \\
  \label{eq:FourierProperties4}
\mathcal{F}\left(\Pi \right) & = & S \,{\textrm {sinc}}(\pi \sqrt{S} \kappa_x) \,{\textrm {sinc}}(\pi \sqrt{S} \kappa_y) \\
  \label{eq:FourierProperties5}
\mathcal{F}(D_{\phi}) & = & 2\delta_{\textrm{Dirac}}(\V{\kappa}) - 2 \mathcal{S}_{\phi}(\V{\kappa})\, ,
\end{eqnarray}
where $\mathcal{S}_{\phi}(\V{\kappa})$ is the power spectral density of the
phase, in square radians, known as (by Fourier Transform of
Eq.~(\ref{eq:ScaledSF}))
\begin{equation}
 \label{eq:ScaledSpectr}
\mathcal{S}_{\phi}(\V{\kappa}) =  \left\{
          \begin{array}{ll}
            0.0229 \, r_0^{-5/3}\, \Norm{\V{\kappa}}^{-11/3} & \mathrm{if}\quad r_0/L_0 =0 \\
            0.0229 \, r_0^{-5/3} \left( \Norm{\V{\kappa}}^2 + \frac {1} {L_0^2} \right)^{-11/6} ~ & \mathrm{otherwise} \\
          \end{array}
        \right.
\end{equation}

Then Eqs.~(\ref{eq:SxxVidal}-\ref{eq:SyyVidal}) can be simplified into
\begin{eqnarray}
  \label{eq:SxxRed}
  \mathcal{S}_{Sxx}(\V{\kappa}) = & 4\, \pi^2 \kappa_x^2 & \, {\textrm {sinc}}^2(\pi \sqrt{S} \kappa_x) \,{\textrm {sinc}}^2(\pi \sqrt{S} \kappa_y) \,\, \mathcal{S}_{\phi}(\V{\kappa})\\
\label{eq:SxyRed}
  \mathcal{S}_{Sxy}(\V{\kappa}) = & 4\, \pi^2 \kappa_x \kappa_y & \, {\textrm {sinc}}^2(\pi \sqrt{S} \kappa_x) \,{\textrm {sinc}}^2(\pi \sqrt{S} \kappa_y) \,\, \mathcal{S}_{\phi}(\V{\kappa})\\
\label{eq:SyyRed}
  \mathcal{S}_{Syy}(\V{\kappa}) = & 4\, \pi^2 \kappa_y^2 & \, {\textrm {sinc}}^2(\pi \sqrt{S} \kappa_x) \,{\textrm {sinc}}^2(\pi \sqrt{S} \kappa_y) \,\, \mathcal{S}_{\phi}(\V{\kappa})\,.
\end{eqnarray}
These simple expressions are obtained thanks to the fact that the Dirac part
of Eq.~(\ref{eq:FourierProperties5}) is never non zero at the same time as
the factors $\kappa_x^2$, $\kappa_x\,\kappa_y$ and $\kappa_y^2$ in Eqs.~(\ref{eq:SxxRed})-(\ref{eq:SyyRed})\,.

\section{Fourier-Domain covariance of incremental Shack-Hartmann
  slopes}
\label{sec:DiffFourierCovMap}

In the same way we derived slopes covariance maps in the
Fourier-Domain in Appendix~\ref{sec:CovS-Fourier}, we can express the
incremental slopes covariance maps in the Fourier-Domain. In a first
step, let consider a single turbulent layer translating horizontally
in the pupil plane with wind speed vector $\V{V}$, and let us note
$\tau_{e}$ the WFS exposure time. Taylor frozen flow assumption is
expressed here as
\begin{equation}
  \label{eq:TaylorWF}
  \delta \V{\phi}_k(\V{r}) = \V{\phi}_{k+1}(\V{r}) - \V{\phi}_k(\V{r}) = \V{\phi}_k(\V{r}-\V{V}\,\tau_e) - \V{\phi}_k(\V{r})\,.
\end{equation}

It is then possible to use similar formulae as
Eqs.~(\ref{eq:SxxVidal})-(\ref{eq:SyyVidal}) but in terms of covariance
incremental slopes. Measuring the slope at a posterior instant $\V{V}
\tau_e$ is equivalent to translate the subaperture center coordinates by
$-\V{V}\tau_e$. As a consequence,
\begin{equation}
  \label{eq:SdeltasxxDev}
  \Avg{\delta s_x(0) \cdot \delta s_x(\V{r})} = 2\, \Avg{s_x(0)\cdot s_x(\V{r})} - \Avg{s_x(0)\cdot s_x(\V{r}-\V{V}\tau_e)} - \Avg{s_x(0)\cdot s_x(\V{r}+\V{V}\tau_e)}\,.
\end{equation}

Using the fact that a translation of $\V{V}\tau_e$ in $(x,y)-$space is
equivalent to a multiplication by $\exp(j \V{\kappa} \cdot\V{V} \tau_e)$ in Fourier
Domain, then
\begin{eqnarray}
  \label{eq:SdeltasxxRed}
  \mathcal{S}_{\delta Sxx}(\V{\kappa}) & = & 4 \,\mathcal{S}_{Sxx}(\V{\kappa})  \, \sin^2 (\pi \V{\kappa} \cdot \V{V}\tau_e) \\
\label{eq:SdeltasxyRed}
  \mathcal{S}_{\delta Sxy}(\V{\kappa}) & = & 4 \,\mathcal{S}_{Sxy}(\V{\kappa})  \, \sin^2 (\pi \V{\kappa} \cdot \V{V}\tau_e) \\
\label{eq:SdeltasyyRed}
  \mathcal{S}_{\delta Syy}(\V{\kappa}) & = & 4 \,\mathcal{S}_{Syy}(\V{\kappa})  \, \sin^2 (\pi \V{\kappa} \cdot \V{V}\tau_e) \,,  
\end{eqnarray}
where $ \mathcal{S}_{Sxx}(\V{\kappa})$, $ \mathcal{S}_{Sxy}(\V{\kappa})$
and $ \mathcal{S}_{Syy}(\V{\kappa})$ are given by
Eqs.~(\ref{eq:SxxRed})-(\ref{eq:SyyRed}). If the wind speed is zero, then
obviously the incremental slopes are zero everywhere. The way the
Fourier-Domain slopes covariance are weighted by the sine squared in the
expression of the Fourier-Domain incremental covariances depends on the
wind speed direction $\V{V}$. It is a zero factor for frequencies
orthogonal to the wind speed, and it increases along with the frequency
component collinear to the wind, $\V{\kappa}\cdot \V{V}$.

The atmosphere can be considered to have several turbulent layers
(statistically independent), so that the sum of the contribution of
each layer is used to compute the spectral density of the incremental
slopes. Assuming $n_l$ layers with the turbulence strength ratio
${C_n^2}_{i_l}$, and wind speed $\V{V}_{i_l}$ standing for the
$i_l-$th layer and the global atmosphere Fried parameter noted $r_0$,
we can write
\begin{equation}
  \label{eq:GlobalSdeltasxx}  
    \mathcal{S}_{\delta Sxx}(\V{\kappa}) = 4 \,\mathcal{S}_{Sxx}(\V{\kappa})  \, \sum_{i_l=1}^{n_l} ({C_n^2}_{i_l} \, \sin^2 (\pi \V{\kappa} \cdot \V{V}_{i_l}\tau_e) )\,,
\end{equation}
and the same sum factor applies to $\mathcal{S}_{\delta Sxy}(\V{\kappa})$
and $\mathcal{S}_{\delta Syy}(\V{\kappa})$ in
Eqs.~(\ref{eq:SdeltasxyRed})-(\ref{eq:SdeltasyyRed}). Due to the fact that
the wind speed distribution in altitude in the turbulent atmosphere may
present various directions, the total sum factor in
Eq.~(\ref{eq:GlobalSdeltasxx}) is not in general radially symmetric.

\section{Equivalence of implementations}
\label{sec:equivKolb}

We demonstrate how to go from Eq.~(\ref{eq:solGinvert}) to
Eq.~(\ref{eq:GJKO}). From Eq.~(\ref{eq:solGinvert}), the $(i,j)-$th
element of matrix $\M{G}^{*}$ can be decomposed as follows:
\begin{eqnarray}
  \label{eq:GCBEdecomp}
  \M{G}^{*}(i,j) & = &- \sum_{k=1}^{n_a} \M{C}^*_{\delta d, \delta a}(i,k)\, \M{C}^{*\, \dag}_{\delta a, \delta a} (k,j)\\
& = & - \frac{1} {N}\, \sum_{k=1}^{n_a} \sum_{l=1}^{N} \V{\delta d}(i,l) \, \V{\delta a}(k,l) \, \M{C}^{*\, \dag}_{\delta a, \delta a} (k,j)\\
  \label{eq:GCBEdecompCdada}
& = & - \frac{1} {N}\, \sum_{l=1}^{N} \V{\delta d}(i,l) \, \left[ \sum_{k=1}^{n_a} \V{\delta a}(k,l) \, \M{C}^{*\, \dag}_{\delta a, \delta a} (k,j) \right]
\end{eqnarray}

In addition, by definition,
\begin{equation}
  \label{eq:DefCdada}
  \M{C}^{*}_{\delta a, \delta a} = \frac {1} {N} \,({\Delta_a} \cdot {\Delta_a}^{\T})\,,
\end{equation}
so that Eq.~(\ref{eq:GCBEdecompCdada}) becomes
\begin{eqnarray}
   \M{G}^{*}(i,j) & = &- \sum_{l=1}^{N} \V{\delta d}(i,l) \, \sum_{k=1}^{n_a} \V{\delta a}(k,l) \,({\Delta_a} \cdot {\Delta_a}^{\T})^{\dag}(k,j)\,.\\
 \label{eq:GCBEDeltaDelta}
& = &- \sum_{l=1}^{N} \V{\delta d}(i,l) \, \sum_{k=1}^{n_a} {\Delta_a}(k,l) \,({\Delta_a} \cdot {\Delta_a}^{\T})^{\dag}(k,j)\,,
\end{eqnarray}
because, with the chosen notations, ${\Delta_a}(k,l)= \V{\delta a}(k,l)$.

Finally, the pseudo-inverse definition allows us to write
\begin{equation}
  \label{eq:PseudoInvTrick}
 {\Delta_a}^{\dag}= {\Delta_a}^{\T} \cdot ({\Delta_a} \cdot {\Delta_a}^{\T})^{\dag}\,,
\end{equation}
so that Eq.~(\ref{eq:GCBEDeltaDelta}) can also be written 
\begin{eqnarray}
  \label{eq:GCBE2JKO}
  \M{G}^{*}(i,j) & = & - \sum_{l=1}^{N} \V{\delta d}(i,l) \, {\Delta_a}^{\dag}(l,j)\\
  & = & - \sum_{l=1}^{N} \Delta_d(i,l) {\Delta_a}^{\dag}(l,j)\,,
\end{eqnarray}
which means 
that
\begin{equation}
  \label{eq:CBE2JKOdone}
  \M{G}^{*} = - \Delta_d \cdot {\Delta_a}^{\dag}\,,
\end{equation}
as exploited in Eq.~(\ref{eq:GJKO}). We have thus demonstrated the
equivalence between the two formulae~(\ref{eq:solGinvert}) and
(\ref{eq:GJKO}).

\acknowledgments     
 
C. B\'echet particularly thanks Johann Kolb and Pierre-Yves Madec from
the European Southern Observatory (Garching, Germany), for fruitful
discussions about the presented method. This work has been partially
funded by the OPTICON-JRA2 project of the European Commission FP7
programme, under Grant Agreement number 226604.


\bibliography{report}   
\bibliographystyle{spiebib}   

\end{document}